\documentclass[epj,referee]{svjour}
\usepackage{graphics,epsfig}
\begin{document}
\title{Corner wetting in a far-from-equilibrium magnetic growth model}
\author{Virginia Man\'{\i}as\inst{1}, Juli\'an Candia\inst{1,2}
\and Ezequiel V. Albano\inst{3}
}                     
%
%
\institute{IFLP,CONICET/Departamento de F\'{\i}sica, Universidad Nacional de La Plata,
C.C. 67, 1900 La Plata, Argentina \and 
The Abdus Salam International Centre for Theoretical Physics,
Strada Costiera 11, 34014 Trieste, Italy \and
INIFTA, 
Universidad Nacional de La Plata, CONICET, C.C. 16, 
Sucursal 4, 1900 La Plata, Argentina}
\date{Received: } 
%
\abstract{
The irreversible growth of magnetic films is studied in three-dimensional 
confined geometries of size $L\times L\times M$, where $M\gg L$ is the 
growing direction. Competing surface magnetic fields, applied 
to opposite corners of the growing system, 
lead to the observation of a localization-delocalization 
(weakly rounded) transition of the interface between domains of 
up and down spins on the planes transverse to the growing direction.
This effective transition is the precursor of a true  
far-from-equilibrium corner wetting transition that takes place in the
thermodynamic limit. The phenomenon is characterized 
quantitatively by drawing a magnetic field-temperature phase diagram,
firstly for a confined sample of finite size, and then by extrapolating
results, obtained with samples of different size, to 
the thermodynamic limit. The results of this work are a nonequilibrium 
realization of analogous phenomena recently investigated in 
equilibrium systems, such as corner wetting transitions in the Ising model.      
} 
\authorrunning {Virginia Manias et al.} 

\maketitle
\section{Introduction}
\label{intro}
Certainly, the phenomenon of wetting is a problem of primary importance 
in many fields of practical technological applications (such as
lubrication, efficiency of detergents, oil recovery in porous materials, 
stability of paint coatings, interaction of macromolecules with 
interfaces, etc. \cite{deg85}), while it is also an attractive, 
challenging phenomenon from the theoretical point of view. Indeed, 
many theoretical efforts involving different approaches have been
devoted to the study of wetting transitions at interfaces under equilibrium 
conditions \cite{par90,swi91,bin95,mac96a,mac96b,wen97,liu98,fer98,fri99,mue00}
(for general overviews of the subject see e.g. \cite{tel,die}), 
and more recently some investigations on nonequilibrium wetting 
phenomena \cite{hin97,can02a,can02b} have been carried out as well.

A related phenomenon, which has attracted great interest in the past decade, 
is the so-called filling or corner wetting transition of a fluid adsorbed 
on a wedge \cite{che90,nap92,rej99}. At present, it is experimentally feasible 
to control the shape of solid surfaces at a nanoscopic level, and indeed a 
growing variety of new types of interfacial phase transitions have been 
found in fluids confined by such structured substrates \cite{die98}. 
The critical wetting in 2D \cite{par99,par00a,par01a,par02,abr02,ania} and 
3D \cite{par00b,par01b} wedges has also recently been investigated 
theoretically under equilibrium conditions. In particular, it should be 
mentioned that the exact solution for a two-dimensional rectangular Ising 
ferromagnet forming a corner with a surface field applied to the spins 
on edges has recently been obtained \cite{ania}. Furthermore, many 
theoretical  results have later been corroborated by extensive Monte 
Carlo simulations performed in both 2D \cite{alb03} and 3D \cite{andrey}.

Within the context of these recent developments,
the aim of this work is to investigate the phenomenon of corner wetting under
far-from-equilibrium conditions by means of a Monte Carlo approach.
To our best knowledge, this paper presents the first study of 
nonequilibrium wedge wetting. It is also worth mentioning that the
irreversible filling of cavities with magnetic materials is also a frontier
topic in the field of the physics of new materials, since modern trends
in technology \cite{fi1} require the characterization of the filling of 
templates, containing imprinted nanometer/micrometer sized features,
with a depositing material \cite{fi2,fi3}.    

\section{Brief summary of recent studies on corner wetting under equilibrium.}
\label{corner_summ}

\begin{figure}
\centerline{
\epsfxsize=7.5cm
\epsfysize=7.5cm
\epsfbox{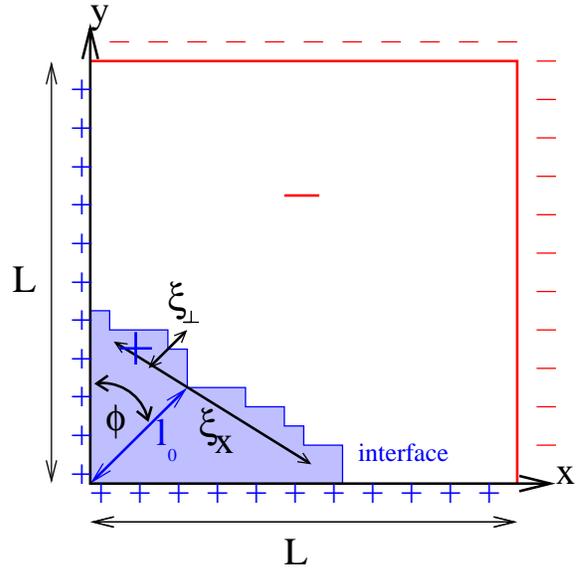}}
\caption{Sketch of the simulation geometry where a transverse
plane ($k = constant$) of size $L \times L$ of the square lattice is shown.
Free boundary conditions are used for the spins in rows $i =1, i = L$
and $j = 1, j = L$, with integers $(i,j) \in [1,L]$ labelling the 
lattice sites. Additionally, a magnetic field $H$ acts on all spins 
in the rows $i = 1$ and $j = 1$, while a magnetic field $-H$ acts 
on all spins in the rows $i = L$ and $j = L$. A typical configuration
of the system below the critical point then contains an interface 
between a domain of positive magnetization $(+)$ and a domain
with negative magnetization $(-)$. Fluctuations of the interface
are characterised by correlation lengths $\xi_{x}$ and $\xi_{\bot}$
in the directions parallel and perpendicular to the interface,
respectively.}  
\label{Fig0}
\end{figure}

Since the present work addresses the phenomena of non-equilibrium corner wetting
in a magnetic material,
it is useful to briefly describe recent progress in the study of their equilibrium
counterpart \cite{par99,par00a,par01a,par02,abr02,ania,par00b,par01b,alb03,andrey}.
Figure \ref{Fig0} shows a sketch of the corner geometry, of size $L \times L$, used 
for the study of corner wetting in two dimensions. Of course, the same sketch 
can be considered as a transverse view of a three dimensional array of size
$L \times L \times M$ (with $M \gg L$), as used in the present work.
 
Let us consider a wedge in contact with the gas phase of a fluid at temperature $T$ 
and chemical potential $\mu$. Based on thermodynamic (macroscopic) arguments \cite{par00a}, 
the theory of corner wetting predicts that the filling of the wedge by the fluid occurs 
at the transition temperature $T_{f}$, given by \cite{par00a,par01a}

\begin{equation}
\Theta_{\pi}(T_{f}) = \pi/2 -  \phi  ,
\label{filling}
\end{equation}

\noindent where $\Theta_{\pi}$ \cite{note} is the contact angle describing 
the droplets on a planar 
interface in the regime of incomplete wetting and  $\phi$ is the angle between
the wall and the diagonal of the wedge (see Figure \ref{Fig0}). Consequently,
$T_{f}$ is smaller than the wetting temperature $T_{w}$ characterising the 
liquid-vapour interaction on a planar substratum.

Similar arguments can be drawn for a magnetic system in a wedge geometry, 
such as the Ising magnet in the square lattice in the presence of competing 
confinement fields as shown in Figure \ref{Fig0}. For the sake of simplicity, let us 
discuss the filling transition (also known as corner-wetting transition) 
in $d = 2$ considering the case  $\phi = \pi/4$ only.
The transition is of second-order and occurs at a critical field $H_{cw}$ \cite{ania}, 
which depends on $T$, and it is observed at temperatures below the order-disorder Onsager  
critical temperature of the Ising magnet. At low temperatures, the interface remains
bounded to one wall and its mean distance to the corner ($ \langle l_{o} \rangle$,
see Figure \ref{Fig0}) diverges according to a power-law behavior when approaching the 
transition. Of course, the growth of the correlation lengths is bound by the finite
lattice size and true divergences are only possible in the thermodynamic limit.
So, one has

\begin{equation}
\langle l_{o} \rangle \propto t^{-\beta_{s}} ,\qquad   t = H_{cw}(T) - H, 
\label{power}
\end{equation}

\noindent where $\beta_{s} = 1$ is the order parameter critical exponent. 
The interface also
develops correlations along the directions parallel and perpendicular to 
it (see Figure \ref{Fig0}). The corresponding correlation lengths 
($\xi_{x}$ and $\xi_{\bot}$, respectively), also diverge at the critical 
point according to

\begin{equation}
\xi_{x} \propto t^{-\nu_{x}} , \qquad  \xi_{\bot} \propto t^{-\nu_{\bot}} ,  
\label{correl}
\end{equation}
    
\noindent where the correlation length exponents are $\nu_{x} = 1$ and $\nu_{\bot} = 1$,
respectively.

On the other hand, taking first the limit $L \rightarrow \infty$ and then the limit
$t \rightarrow 0^{+}$, the probability distribution of the mean position of the
interface is expected to decay exponentially, namely

\begin{equation}
P(\langle l_{o} \rangle) = 
\frac{1}{\langle l_{o} \rangle} \exp{-( \frac{ l_{o}}{\langle l_{o} \rangle})}
, \qquad   t \rightarrow 0^{+} . 
\label{distrilo}
\end{equation}
 
\noindent  For finite systems of side $L$, equation (\ref{distrilo}) needs 
to be symmetrized 
with respect to both corners to which the interface can be bound \cite{alb03}. Equation 
(\ref{distrilo}) holds within the regime of incomplete corner wetting
(or equivalently to the case of incomplete corner filling, when the 
interface of the magnetic domain filling the sample is still bound to one corner 
far from the diagonal crossing the sample from left-top to right-bottom (\ref{Fig0}) ). 
However, in the regime of complete corner wetting
(or equivalently complete filling), $t < 0$, the average location 
of the interface lies along the 
diagonal taken from the upper-left corner  to the lower right corner. So, one has 
$\langle l_{o} \rangle = L/\sqrt{2}$ and the probability distribution becomes

\begin{equation}
P(\langle l_{o} \rangle) \propto 
\exp{-( \frac{ [l_{o} - \langle l_{o} \rangle]^{2} } {2 \xi_{\bot}^{2}}) } , 
\label{distri}
\end{equation}
    
\noindent with  \cite{par01a}

\begin{equation}
\xi_{\bot} \propto L^{\frac{1}{2}} ,
\end{equation}

\noindent since it can be interpreted in terms of a random walk description of the 
interface in $d = 2$, such that excursions in the direction perpendicular to the 
average interface orientation add up randomly \cite{fish}. 
 
Equation (\ref{distri}) implies that for complete wetting, the average 
magnetization vanishes ($\langle m \rangle = 0$) and the Gaussian distribution of the 
interface location translates into a Gaussian distribution of 
the magnetization given by

\begin{equation}
{P}(m) \sim \exp\left(-\frac{m^2\,L^2}{2k_BT\,\chi_L}\right) ,
\label{magne}
\end{equation}
\noindent with 
\begin{equation}
\chi_L \propto \xi_{\bot}^{2} \propto   L,
\label{chicho}
\end{equation}

\noindent where the size-dependent width of the Gaussian is given
by $\chi_{L} \equiv ( \langle m^2 \rangle - \langle m \rangle^2 )$,
with $\langle m \rangle^2 = 0$ within the complete wetting regime.    
It should be noticed that, as in the case of the localization-delocalization 
transition observed for the Ising magnet in a slit geometry \cite{par90,alb89}, in 
the case of complete wetting the interface is no longer bound to 
the wall (or corner, respectively).

\section{The Magnetic Eden Model in a corner geometry and the 
simulation method}
\label{mem_simu}
For definiteness, we will adopt a magnetic language throughout, although 
the relevant physical concepts discussed here can be extended to other 
systems such as fluids, polymers, and binary mixtures.
The irreversible growth of a ferromagnetic material has been studied by 
using the so-called magnetic Eden model (MEM) \cite{aus93}, an extension 
of the classical Eden model \cite{ede58} in which the growing particles 
have an additional degree of freedom, i.e., the spin.
Several recent investigations based on this model showed a rich variety 
of interesting phenomena, such as the occurrence of morphological 
phase transitions associated with the growing interface \cite{can00},
Ising-like order-disorder phase transitions \cite{can01a}, spontaneous 
magnetization reversals \cite{can01b}, and a far-from-equilibrium wetting 
transition driven by competing surface magnetic fields \cite{can02a,can02b}.

The MEM in $(2 + 1)-$dimensions is studied in the square lattice 
by using a rectangular geometry $L \times L \times M$ (with $M \gg L$).
The location of each spin on the lattice is specified through its 
coordinates $(i,j,k)$, ($1 \leq i,j \leq L$, $1 \leq k \leq M$). 
A transverse plane of the used geometry, taken for $k = constant$, 
is sketched in Figure \ref{Fig0}. 
The starting seed for the growing cluster is a plane of $L \times L$ 
parallel-oriented spins placed at $k=1$, and cluster growth takes place 
along the positive longitudinal direction (i.e., $k \geq  2$).
Open boundary conditions are adopted in both transverse
directions, and competing surface magnetic fields  $H>0$ and $H'=-H$
are applied to opposite corners, as shown in Figure \ref{Fig0}. 
Then, clusters are grown by selectively adding
spins ($S_{ijk}= \pm 1$) to perimeter sites, which are defined as the
nearest-neighbour (NN) empty sites of the already occupied ones.
Considering a ferromagnetic interaction of strength $J>0$ between NN spins, 
the energy $E$ of a given configuration of spins is given by 

\begin{eqnarray}
E=-\frac{J}{2}\left(\sum_{\langle ijk,i^{'}j^{'}k^{'}\rangle}
S_{ijk}S_{i^{'}j^{'}k^{'}}\right) \nonumber\\
-H\left(\sum_{\langle jk\rangle} S_{1jk}+
\sum_{\langle ik\rangle }S_{i1k}\right)
+H\left(
\sum_{\langle jk\rangle} S_{Ljk}+
\sum_{\langle ik\rangle }S_{iLk}\right)\ ,
\end{eqnarray}
\noindent where the summation $\langle ijk,i^{'}j^{'}k^{'}\rangle$ 
is taken over occupied NN sites, while the remaining terms 
are sums over occupied surface sites, in order to 
take into account the effect of the surface magnetic fields.  
The Boltzmann constant is set equal to unity throughout, 
and the temperature, energy, and magnetic fields are measured 
in units of $J$. The probability for a perimeter site
to be occupied by a spin is taken as proportional to the Boltzmann factor
$\exp(- \frac{\Delta E}{T})$, where $\Delta E$
is the change of energy involved in the addition of
the spin. At each step, the probabilities of adding up and down
spins to a given site have to be evaluated for all perimeter sites.
After proper normalization of the probabilities, the growing site and 
the orientation of the spin are determined with Monte Carlo techniques.
For additional details on the MEM and the simulation method
see e.g. \cite{can02a,can02b,aus93,can00,can01b}.

Since the observables of interest (e.g. the mean transverse magnetization 
and its fluctuations, see below) require the growth of samples with 
a large number of transverse planes of size $L\times L$,
clusters having up to $10^9$ spins have typically been grown for lattice 
sizes in the range $12\leq L\leq 96$. It should be mentioned that
the growing front leaves voids behind that are incorporated to the bulk
of the sample during some  transient period. However, 
since these voids are also perimeter sites they are ultimately filled 
in during the growth process. Hence, far behind the active growth 
interface, the system is compact and frozen. So, the different 
quantities of interest are measured on defect-free transverse planes.

Moreover, the update algorithm is quite
time consuming, as compared to standard Ising simulations, since the growing 
probability has to be computed after each deposition event. Hence, it should 
be noticed that the results obtained in this work involve a large 
computational effort. Let us also remark that, although both the interaction 
energy and the Boltzmann probability distribution considered for the MEM are
similar to those used for the Ising model with surface magnetic fields,
these two models operate under extremely different conditions, since the MEM
describes the irreversible growth of a magnetic material, while
the Ising model deals with a magnet under equilibrium.
Previous studies have demonstrated that the MEM in $(1+1)-$dimensions
is not critical but exhibits a second-order transition
at $T_c=0.69\pm 0.01$ in $(2+1)-$dimensions \cite{can01a}.

\section{Results and discussion}
\label{res_disc}

The growth of magnetic Eden aggregates in a $(d + 1)-$confined geometry is 
characterized by an initial transient period followed by a 
nonequilibrium stationary state that is independent of the initial
configuration \cite{can02a}. The length of the transient is of 
the order of $L^{d}$ (d = 2 in the present study), and the 
proportionality constant depends on $T$. 
We have checked that disregarding the first $100 L^{2}$ 
transverse planes, subsequent averages are independent of
the initial configuration. So, all results reported in this paper 
are obtained under these conditions.

\begin{figure}
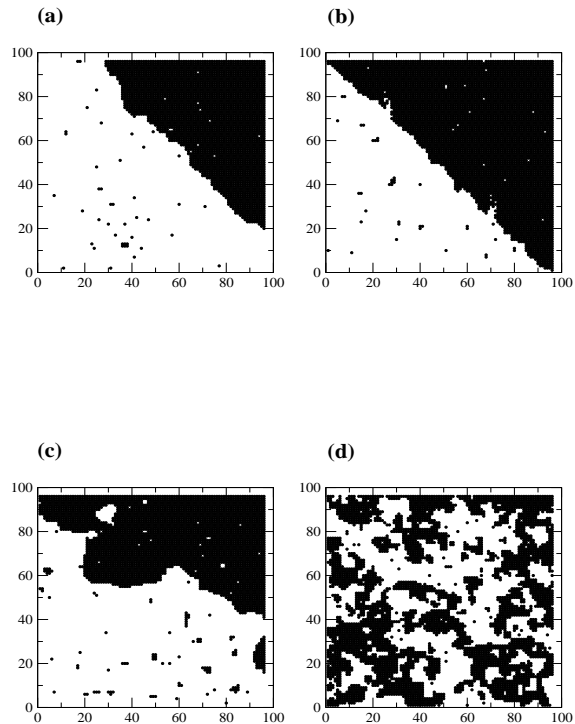

\centerline{
\epsfxsize=7.5cm
\epsfysize=3.75cm
\vspace*{2.0cm}       
\epsfbox{Fig2ab.eps}}
\centerline{
\epsfxsize=7.5cm
\epsfysize=3.75cm
\vspace*{1cm}       
\epsfbox{Fig2cd.eps}}
\caption{Typical snapshot configurations of transverse ($k=constant$) planes. 
Black points correspond to down spins, while up spins are left in white. 
The surface field is positive (negative) on the left and bottom (right and top)  
sides of the wedge, as shown in Figure \ref{Fig0}.
The snapshots correspond to lattices of size $L=96$ and are obtained 
keeping the surface fields constant and varying the temperature, as follows: 
(a) $H=0.4,\ T=0.6$; (b) $H=0.4,\ T=0.65$, (c) $H=0.4,\ T=0.7$, 
and (d) $H=0.4,\ T=1.0$.}  
\label{Fig1}
\end{figure}

\begin{figure}
\centerline{
\epsfxsize=7.5cm
\epsfysize=6.5cm
\epsfbox{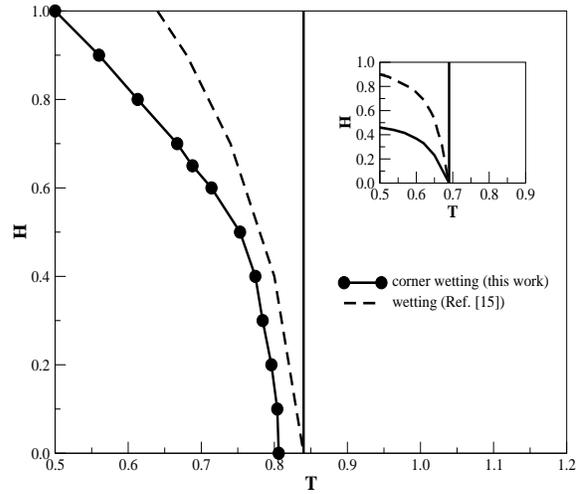}}
\caption{The $H-T$ phase diagram corresponding to a lattice of size $L=12$.
The vertical straight line at $T_c(L)=0.84$ shows the $L-$dependent critical 
temperature of the MEM in the absence of any magnetic field,
which separates the low-temperature ordered phase from the 
high-temperature disordered phase.
Besides, the corner localization-delocalization transition 
curve (full circles and solid line) 
obtained in this work, as well as the localization-delocalization 
transition curve (dashed line) 
that corresponds to the MEM with competing surface magnetic fields
applied to parallel confinement walls \cite{can02a}, is shown for comparison. 
The inset shows the phase diagram corresponding to the thermodynamic limit, 
as obtained by extrapolating results obtained for systems of different 
lattice size. Again, the corner wetting transition (solid line)
is compared to the wetting transition of the MEM (dashed line) obtained in 
Reference \cite{can02a}. More details in the text.} 
\label{Fig2}
\end{figure}

Depending on the temperature and the surface magnetic field considered, 
different regimes are clearly distinguished upon growth of a 
magnetic aggregate in the confined corner geometry, as illustrated by 
the typical snapshot configurations of transverse ($k=constant$) planes 
shown in Figure \ref{Fig1}. Furthermore, in order to characterize 
these different phases quantitatively, 
we will locate the boundaries between them in a magnetic field {\it versus} 
temperature ($H-T$) phase diagram (see Figure \ref{Fig2}). 
Below, we will first describe the results for a confined, finite-size system, 
and then we will obtain the phase diagram in the thermodynamic 
limit by means of extrapolation procedures.

Let us now discuss the localization-delocalization behavior of the interface
observed when the value of the surface field is fixed ($H = 0.4$)
but the temperature is increased. 
For this qualitative description we will use the typical
snapshot configurations shown in Figure \ref{Fig1}.  
Figure \ref{Fig1}(a) shows the regime characteristic of low temperatures
(i.e. far below the critical temperature of the MEM). Here, the interface 
between up and down domains is localized close to the upper-right corner. 
Of course, during very long-time simulations one  observes the localization of 
the interface close to both corners with the same probability. Keeping the same 
field but increasing the temperature, the interface starts to depart
from the corners and, close to the effective corner wetting transition 
temperature of the considered lattice, it remains still localized close to the 
diagonal of the sample, as shown in Figure \ref{Fig1}(b). 
However, further increase of the temperature causes the interface 
to become delocalized (see Figure \ref{Fig1}(c)). Here the domain of positive 
spins makes frequent excursions up to the boundary with negative field (see the 
lower-right corner of Figure \ref{Fig1}(c)). Also, notice that a relatively 
weaker excursion of the negative domain to the positive boundary field 
is observed at the upper-left corner. Of course, due to the symmetry of the 
system, these excursions cause the temporal average position of the interface 
to be just at the diagonal of the sample with $\langle l_{o} \rangle = L/\sqrt{2}$,
while its probability distribution is given by equation (\ref{distri}). 
For this regime the total magnetization vanishes  and its probability distribution
is a Gaussian centered around $ m  = 0$, see equation (\ref{magne}).   
Finally, at higher temperatures and within the disordered phase (i.e., above the
critical temperature of the MEM), the interface between domains of 
up and down spins vanishes due to the thermal noise, as shown by the 
typical high-temperature configuration of Figure \ref{Fig1}(d). 

Let us now discuss the interface localization-delocalization 
``phase diagram'' of a confined  system of lateral size $L = 12$, 
as shown in Figure \ref{Fig2}. 
\begin{figure}
\centerline{
\epsfxsize=7.5cm
\epsfysize=6.5cm
\epsfbox{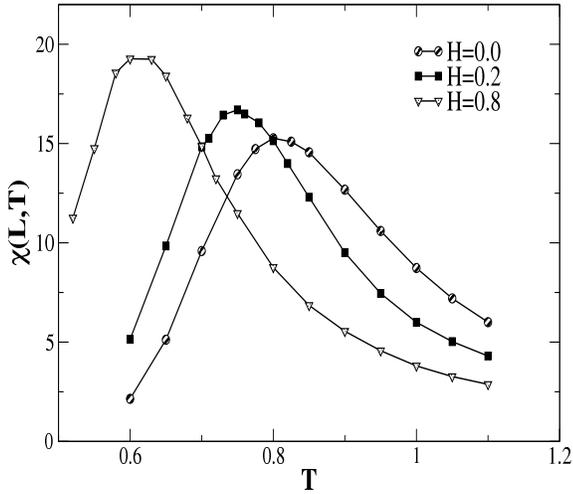}}
\caption{Plots of the magnetization fluctuations as a function 
of the temperature obtained for different values of the magnetic field, 
as indicated. Following a 
standard procedure \cite{alb89}, the maxima of $\chi(L,T)$ define the points of 
the corner localization-delocalization transition curve on the $H-T$ plane, as 
shown in Figure \ref{Fig2}.}
\label{Fig3} 
\end{figure}

In order to locate the transition curve we have measured the 
magnetization fluctuations, given by
\begin{equation}
\chi(L,T) = \frac{L^2}{T}\left(\langle m^2\rangle-\langle|m|\rangle^2\right)\  ,
\label{chimi}
\end{equation}
\noindent where $\langle ...\rangle$ means the average taken over a large 
number of transverse planes in the stationary regime, and $m=m(k,L,T,H)$ 
is the mean transverse magnetization given by
\begin{equation}
m=\frac{1}{L^2}\sum_{i,j=1}^LS_{ijk}\ .
\label{magneti}
\end{equation}

It should be noticed that under equilibrium conditions and provided 
that the fluctuation-dissipation theorem holds, the fluctuations of the 
magnetization can be identified with the susceptibility. However,
this may not be the case for the present far-from-equilibrium system.

Working under equilibrium \cite{bindbook} and far from 
it \cite{can01b,novo1,novo2}
is usual to locate $L-$dependent 'critical' points at the maximum 
of $\chi(L,T)$. This procedure has proved to be also useful to locate 
$L-$dependent localization-delocalization transition points (namely
wetting ``critical'' temperatures $T_{cw}(L)$ )
under
equilibrium \cite{alb89} and far from it \cite{can02a}. So, we have also 
used the same method in the present case. In order to 
illustrate this procedure, Figure \ref{Fig3} shows plots of $\chi(L,T)$ 
as a function of the temperature for different values of the surface magnetic 
field. The maximum of each curve defines a point in the 
transition curve shown in Figure \ref{Fig2}. It should be noticed that 
the localization-delocalization transition in a confined system 
actually corresponds to a so-called quasi-corner wetting transition, 
which is indeed the precursor of the true corner wetting transition that 
occurs in the thermodynamic limit.

In Figure \ref{Fig2}, the vertical straight 
line at $T_c(L=12) = 0.84$ shows the location of the 
$L-$dependent ``critical'' temperature of the finite system in the absence
of surface fields. This value was obtained previously for the MEM in the absence 
of an external magnetic field and corresponds to the slit geometry \cite{can01a}.
Although, as it is well known from finite-size scaling theory, 
there is some degree of 
arbitrariness in locating the $L-$dependent critical temperature $T_c(L)$ 
of a finite system, the actual critical point $T_c$ of the infinite system, 
which can be obtained by extrapolating $T_c(L)$ to the $L\to\infty$ limit, 
is unique and independent of any particular choice for the finite-size 
critical point. Let us also recall that for low temperatures ($T<T_c(L)$) 
the phase diagram 
corresponds to the ordered growth regime, while high temperatures 
($T\geq T_c(L)$) are associated with the disordered growth regime. 

For the sake of comparison, the localization-delocalization transition curve 
obtained in this work for the corner geometry is compared to the 
localization-delocalization transition curve of the MEM
(dashed line in Figure \ref{Fig2}) that corresponds to competing surface 
magnetic fields applied to parallel confinement walls \cite{can02a}, 
i.e., the so-called slit geometry. Notice that in this case the 
critical curve ends, for $H = 0$, at the $L-$dependent critical 
point  $T_{c}(L=12) = 0.84$.
It is observed that the phase diagrams obtained here are similar to the analogous 
phenomena of wetting and corner wetting observed in 2D 
equilibrium systems, as e.g. in the confined Ising model \cite{alb03,alb89}.
In fact, for a given surface field, in the case of corner wetting, 
the corresponding critical temperature is always smaller  than for pure wetting
except, of course, for $H = 0$. This scenario is in agreement with 
equation (\ref{filling}), which was obtained on the basis of thermodynamic 
(equilibrium) considerations \cite{par00a}.

It should be noticed that the transition curve corresponding to the 
corner geometry intercepts the 
horizontal axis $(H=0)$ close to $T_c(L=12)\simeq 0.81$, i.e.,
a value slightly smaller than the $L-$dependent order-disorder
critical temperature of the MEM model in the absence of surface fields.
This shift simply reflects the different boundary conditions
that one has to use to study both systems. In fact, for the corner 
geometry one has to apply open boundary conditions, while for studying the 
confinement due to parallel walls one has to adopt open (periodic)
boundary conditions along (in the direction perpendicular to) the walls. 
Of course, both localization-delocalization transitions lie within the
ordered growth regime where an interface between domains is well defined.

\begin{figure}
\centerline{
\epsfxsize=7.5cm
\epsfysize=7.5cm
\epsfbox{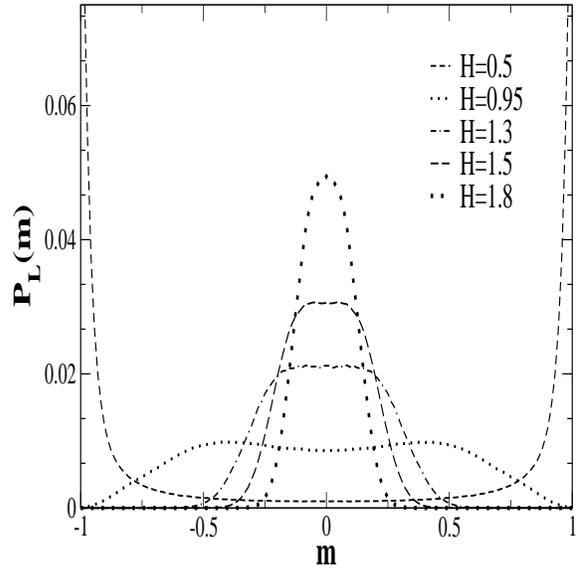}}
\caption{Order parameter probability distributions obtained for 
$T=0.65$ and different values of the field, as indicated. 
The lattice size is $L = 12$.}
\label{Fig4}  
\end{figure}

\begin{figure}
\centerline{
\epsfxsize=7.5cm
\epsfysize=7.5cm
\epsfbox{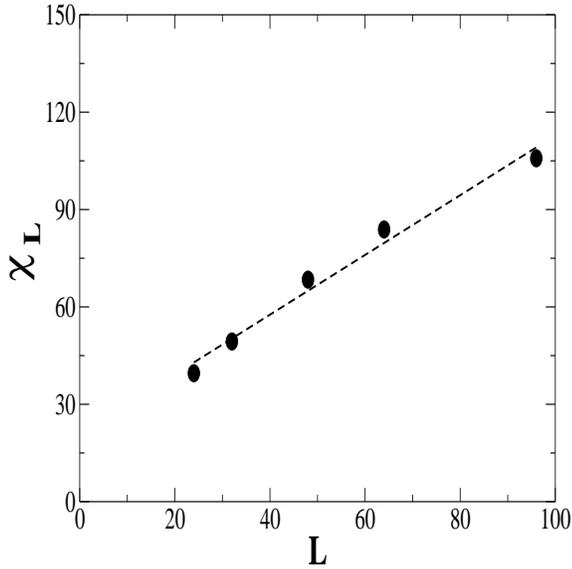}}
\caption{Plot of $\chi_{L}$ versus $L$, as obtained from the order
parameter probability distribution measured within the wet phase,
for $H = 0.8$ and $T = 0.65$.} 
\label{Fig5}
\end{figure}

Further insights can be gained by examining the behavior of the probability 
distribution of the mean transverse magnetization, $P_L(m)$. In fact, in 
the thermodynamic limit, the order parameter probability distribution of 
an equilibrium system at criticality is universal (up to rescaling of the 
order parameter), and hence it contains information on the universality
class of the system \cite{bin81,bru81,tsy00}.  
Figure \ref{Fig4} shows the dependence of $P_L(m)$ on the surface magnetic
field  for fixed values of the temperature ($T=0.65$) and the lattice size ($L=12$). 
For larger fields ($H = 1.8$ in Figure \ref{Fig4})  
the distribution is a Gaussian centered at $m=0$,
in agreement with the fact that the {\bf average} 
location of the interface (for temperatures below $T_c(L)$)
is a straight line crossing the sample from one corner, where the
magnetic fields have opposite direction, to the other (see e.g. the 
snapshots shown in Figures \ref{Fig1} (b) and (c), as well as the 
corresponding discussion). Hence, the 
average magnetization $\langle m \rangle = 0$, and the Gaussian distribution of the 
interface location given by equation (\ref{distri}) translates into a Gaussian 
distribution of the magnetization given by equation (\ref{magne}).
So, within this wet regime the width of the Gaussian can easily be measured and 
the relationship $\chi_L \propto L$ (see equations (\ref{magne}) and (\ref{chicho}) )
has been verified, as shown in Figure \ref{Fig5}.
This behavior is similar to that observed in the interface 
localization-delocalization transition under equilibrium
conditions, in both the slit \cite{par90,alb89} 
and corner geometries \cite{alb03}.

Decreasing the magnitude of the field, 
the onset of two distinct maxima shows 
gradually up (e.g. for $H = 1.8$ in Figure \ref{Fig4}), 
as is also observed in equilibrium 
systems \cite{par90,alb03,andrey,alb89}. 
Indeed, this behavior not only reflects the fact that the interface is located
close to one of both corners with the same probability, but also it is 
the signature of a thermal continuous phase transition taking 
place at finite critical temperature \cite{bin81,bru81,tsy00}.
Finally, well inside the nonwet phase, for $H = 0.5$ in Figure \ref{Fig4}),
one has an exponential decay of the distribution as already reported
for the equilibrium counterpart \cite{alb03}.   

\begin{figure}
\centerline{
\epsfxsize=7.5cm
\epsfysize=7.5cm
\epsfbox{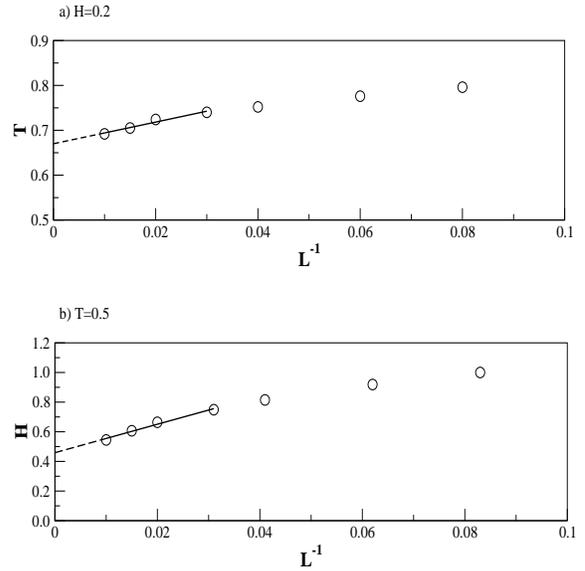}}
\caption{Extrapolation of the results to the thermodynamic limit: 
(a) plot of $T_{cw}(L)$ versus $L^{-1}$, for $H_{cw}=0.2$, and
(b) plot of $H_{cw}(L)$ versus $L^{-1}$, for $T_{cw}=0.5$.  
The extrapolated fits (which take into account the larger lattices only, 
i.e., $32\leq L\leq 96$) are also shown.} 
\label{Fig6}
\end{figure}

Let us now consider the extrapolation of the $L$-dependent 
critical points to the $L\to\infty$ limit, in order to show that the above 
discussed behavior of the interface is not only characteristic of confined 
systems, but it is also present in the thermodynamic limit, hence leading 
to the phase diagram shown in the inset of Figure \ref{Fig2}.
It is well known that, in the general context of continuous phase transitions,
due to finite-size effects the effective location 
of second-order transitions becomes shifted according to \cite{bindbook}

\begin{equation}
T_{c}(L)  = T_{c}(\infty) + A L^{-1/\nu},
\label{scalT}
\end{equation}

\noindent where $T_{c}(L)$ is the effective $L-$dependent ``critical'' temperature,
$T_{c}(\infty)$ is the actual critical temperature in the thermodynamic limit,
$A$ is a constant and $\nu$ is the correlation length exponent related to the 
proper order parameter. In the present case, the location of the interface
($\langle l_{o} \rangle$), see equation (\ref{power}), is the appropriated 
order parameter for the corner wetting transition, rather than the magnetization
as in the standard Ising (or MEM) model. The associated correlation length
is given by $\xi_{\bot}$ (see equation (\ref{correl})) 
describing the growth of correlations in the direction perpendicular to the
interface (see Figure \ref{Fig0}). For the confined Ising system
in the corner geometry under equilibrium one has that the
predicted exponent is $\nu_{\bot} = 1$. So, we have performed the 
extrapolations to the thermodynamic
limit by taking $\nu = 1$. Of course, an accurate numerical 
determination of the exponent is beyond our computational capabilities.
 
So, let us now apply these concepts to the corner wetting transition of the MEM.
Considering the $L-$ dependent critical points $(H_{cw}(L)$, $T_{cw}(L))$ 
on the localization-delocalization transition curves corresponding 
to finite systems, we fixed the value of either the magnetic field or the 
temperature, and extrapolated the results to the $L^{-1}\to 0$ limit.  
For the purpose of illustration, Figure \ref{Fig6}(a) shows a plot 
of $T_{cw}(L)$ versus $L^{-1}$ obtained for a fixed value
of the magnetic field ($H_{cw} = 0.2$), while Figure \ref{Fig6}(b) 
shows a plot of $H_{cw}(L)$ versus $L^{-1}$ for a fixed value of
the temperature ($T_{cw} = 0.5$). Considering only larger lattices  
(namely, within the range $32\leq L\leq 96$), the extrapolations allow 
us to determine the  corner wetting transition curve corresponding to 
the thermodynamic ($L\to\infty$) limit, which is shown in Figure \ref{Fig2}.
Deviation from the linear behavior observed for smaller lattices may be due
to finite-size corrections or eventually to the fact that the 
correlation length exponent could slightly depart from unity, as
already discussed above. 
It should be noticed that, as in the case of confined (finite-size) systems, 
the corner wetting transition (lower curve, solid line) is here 
compared to the wetting transition of the MEM (upper curve, dashed line) 
obtained in Reference \cite{can02a}. 
As observed in the confined Ising model \cite{alb03,alb89},
the critical curve in the case of corner wetting lies also below than 
that of pure wetting, except for $H = 0$.

\begin{figure}
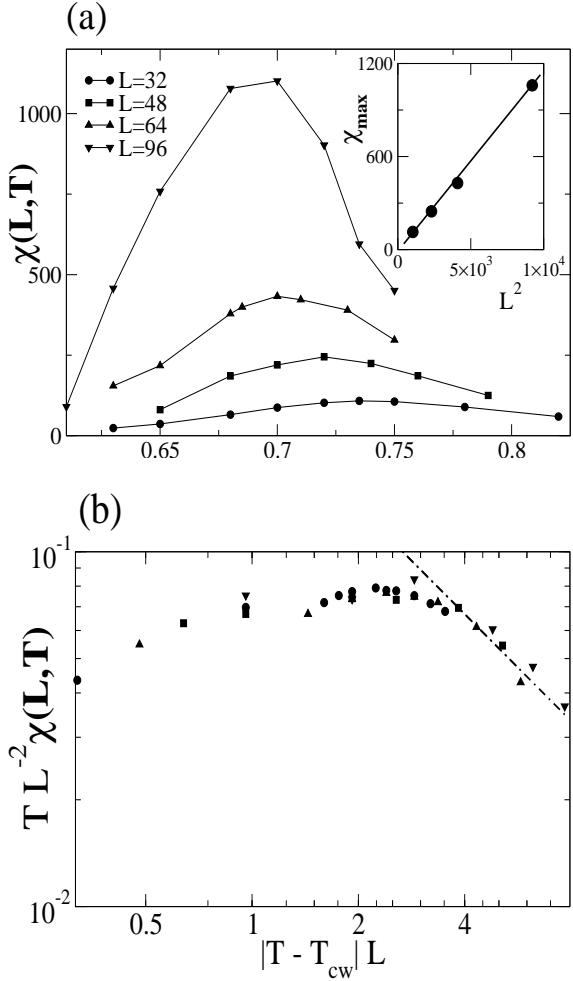

\centerline{
\epsfxsize=7.5cm
\epsfysize=6.5cm
\epsfbox{Fig8a.eps}}
\centerline{
\epsfxsize=7.5cm
\epsfysize=6.5cm
\epsfbox{Fig8b.eps}}
\caption{(a) Plots of the magnetization fluctuations as a function 
of the temperature obtained for different values of the lattice 
size, as indicated. The surface magnetic field has been fixed to $H = 0.2$. 
the inset shows a plot of the maxima of $\chi$ versus $L^2$.
(b) Scaling plot of the data corresponding to the magnetization fluctuations   
shown in (a) performed according to equation (\ref{chima}). The dashed-dotted
line shows the expected asymptotic behavior and has been shown for 
the sake of comparison.}
\label{Fig8} 
\end{figure}

After the evaluation of the critical points it is natural to discuss
the trends of the magnetization fluctuations upon varying the system size 
and, subsequently, analyze whether a scaling plot could be obtained.
Figure \ref{Fig8}(a) shows plots of $\chi(L,T)$ as a function of the temperature 
obtained for different values of $L$ and keeping constant the surface magnetic 
field $H = 0.2$. As expected, the different curves exhibit rounding and 
shifting effects characteristic of finite systems. For the corner wetting 
transition under equilibrium conditions it is known that the 
susceptibility of the finite square scales according to \cite{alb03} 

\begin{equation}
\chi(L,T) \propto \frac{L^2}{T} \chi^*(Lt) ,
\label{chima}
\end{equation}

\noindent where $\chi^*$ is a scaling function and $t = T -T_{cw}$ 
(see also equation (\ref{chimi}) ). So, in the inset of figure \ref{Fig8}(a) 
we show that the maxima of the magnetization fluctuations $\chi_{max}$
(measured at $T_{cw}(L)$ ) of the
non-equilibrium counterpart also scales with $L^2$. Furthermore, figure 
\ref{Fig8}(b) shows a scaling plot of the magnetization fluctuations according to 
equation (\ref{chima}). The data shown correspond to $T > T_{cw}$
and the expected asymptotic behavior (dashed-dotted line) is roughly
achieved for the larger samples only. Of course, a data collapse of better 
quality would be desirable but, regrettably, it is beyond 
our computational capabilities.  
 
\section{Conclusions}

In this work we have studied the phenomenon of corner wetting, observed 
upon the growth of a magnetic material under far-from-equilibrium 
conditions, using extensive Monte Carlo simulations. The occurrence of 
this phenomenon was firstly described at a qualitative level, by means 
of typical snapshot configurations that depend on the 
control parameters (surface magnetic field and temperature), and 
subsequently quantitatively, by identifying well defined regions in 
the $H-T$ phase diagram. After presenting the interface 
localization-delocalization transition phenomenon in (finite-size) 
confined geometries, the results were extrapolated to the thermodynamic limit.
These results, obtained in the framework of nonequilibrium growth systems, 
are a novel realization of analogous phenomena, which have recently been 
observed in equilibrium systems. Hence, we hope that this work 
will contribute to the understanding of the irreversible growth of 
binary mixtures in confined geometries, as well as of wetting-related 
phenomena. 

\subsection*{Acknowledgements}
This work was financially supported by CONICET, UNLP, and ANPCyT (Argentina).
%

\begin{thebibliography}{}
\bibitem{note} The subscript $\pi$ is used to emphasize that this quantity is
pertinent to a planar wetting transition.
\end{thebibliography}
%

\end{document}